\documentclass[journal,twoside,web]{ieeecolor}
\usepackage{generic}
\usepackage{cite}
\usepackage{amsmath,amssymb,amsfonts}
\usepackage{algorithmic}
\usepackage{graphicx}
\usepackage{algorithm,algorithmic}
\usepackage{hyperref}
\usepackage{subcaption}
\hypersetup{hidelinks=true}
\usepackage{textcomp}
\usepackage{caption}
\def\BibTeX{{\rm B\kern-.05em{\sc i\kern-.025em b}\kern-.08em
    T\kern-.1667em\lower.7ex\hbox{E}\kern-.125emX}}
\usepackage{multirow}

\begin{document}
\title{ 
Transparency in Sleep Staging: Deep Learning Method for EEG Sleep Stage 
Classification with Model Interpretability}
\author{Shivam Kumar Sharma\footnotemark*, Suvadeep Maiti\footnotemark*, S. Mythirayee, P R
Srijithesh and Raju Surampudi Bapi
\thanks{\footnotemark* authors contributed equally}
\thanks{Shivm Kumar Sharma, Suvadeep Maiti and Raju Surampudi Bapi are with International Institute of Information Technology Hyderabad, Hyderabad 500032 India  (e-mail: shivam.sharma@research.iiit.ac.in, suvadeep.maiti@research.iiit.ac.in, raju.bapi@iiit.ac.in). }
\thanks{ S. Mythirayee and P.R.Srijithesh are with National Institute of Mental Health and Neurosciences, Bangalore 560029 India  (e-mail: mythugranger@gmail.com, srijitheshpr@gmail.com).}
}

\maketitle

\begin{abstract}
Automated Sleep stage classification using raw single channel EEG is a critical tool for sleep quality assessment and disorder diagnosis. However, modelling the complexity and variability inherent in this signal is a challenging task, limiting their practicality and effectiveness in clinical settings. To mitigate these challenges, this study presents an end-to-end deep learning (DL) model which integrates squeeze and excitation blocks within the residual network to extract features and stacked Bi-LSTM to understand complex temporal dependencies. A distinctive aspect of this study is the adaptation of GradCam for sleep staging, marking the first instance of an explainable DL model in this domain with alignment of its decision-making with sleep expert's insights. We evaluated our model on the publically available datasets (SleepEDF-20, SleepEDF-78, and SHHS), achieving Macro-F1 scores of 82.5, 78.9, and 81.9, respectively. Additionally, a novel training efficiency enhancement strategy was implemented by increasing stride size, leading to 8x faster training times with minimal impact on performance. Comparative analyses underscore our model outperforms all existing baselines, indicating its potential for clinical usage.

\end{abstract}


\section{Introduction}
\label{sec:introduction}
\IEEEPARstart{S}{leep} is a natural, essential physiological process that all humans experience. It is characterized by a state of reduced consciousness, decreased responsiveness to external stimuli, and distinctive changes in brain activity, body posture, and physiological functions. Sleep plays a crucial role in overall health and well-being, as it supports various cognitive functions, physical restoration, and emotional regulation~\cite{1}.

Sleep consists of two primary states: rapid eye movement (REM) sleep and non-rapid eye movement (NREM) sleep. NREM sleep is more intricately divided into three stages: N1, N2, and N3~\cite{2}. N1 represents the lightest stage, N2 serves as a transitional stage, and N3 stands out as the deepest and most restorative stage, often referred to as slow-wave sleep (SWS). The EEG activity corresponding to each of these classes is delineated in Table~\ref{tab:EEGactivity}.

\begin{table}[t]
\centering
\caption{FIVE SLEEP STAGES AND SLEEP MICROSTRUCTURES ACCORDING TO AASM MANUAL}
\label{tab:EEGactivity}
\begin{tabular}{c|c}
\hline
\textbf{Sleep Stage} & \textbf{EEG Activity}                \\ \hline
Wake (W)             & Alpha waves                          \\
NREM 1 (N1)          & Low-amplitude Mixed-frequency (LAMF) \\
NREM 2 (N2)          & Sleep spindles, K-complexes          \\
NREM 3 (N3)          & Delta waves                          \\
REM                  & Low-amplitude Mixed-frequency (LAMF) \\ \hline
\end{tabular}
\end{table}

Sleep stage classification refers to the process of determining the specific sleep stage a person is in during different periods of sleep. This classification is typically done using polysomnography (PSG), which involves monitoring various physiological signals during sleep, such as EEG, EOG, EMG, and ECG~\cite{2,3}. The PSG data is divided into short epochs (usually 30 seconds), and sleep experts manually score each epoch into one of the sleep stages based on established sleep scoring guidelines, such as the Rechtschaffen and Kales (R and K)~\cite{3} or the American Academy of Sleep Medicine (AASM) criteria~\cite{2}.

However, manual classification is labor-intensive and time-consuming. As a result, there has been a growing interest in developing automatic sleep stage classification methods using machine learning techniques, particularly deep learning. Deep learning models can be trained on large datasets of manually scored sleep data to learn patterns and features indicative of different sleep stages. These models can then automatically classify new PSG data into sleep stages with high accuracy, offering a more efficient and scalable alternative to manual scoring~\cite{4}.
Recent advances in deep learning have shown immense potential in automating the sleep stage classification process. Deep learning models, particularly Convolutional Neural Networks (CNNs) \cite{5} and Recurrent Neural Networks (RNNs)~\cite{6,7,8} have proven to be powerful tools for analyzing complex and high-dimensional data, such as EEG signals. CNNs excel at capturing spatial patterns in data, making them suitable for extracting features from raw EEG signals. On the other hand, RNNs are adept at modeling temporal dependencies, a crucial aspect of understanding sequential patterns in sleep EEG data. Combining CNNs and RNNs in hybrid architectures has led to remarkable results, taking advantage of their complementary strengths.


Due to the high variability of EEG signals between individuals and sessions, and their intricate patterns, a dynamic mechanism is necessary to capture these complex details, which elude simple feature extractors. Consequently, we employed an attention-based feature extractor~\cite{10} combined with a temporal context module to address these challenges effectively. Our research paper makes the following contributions:
\begin{enumerate}
\itemsep=0pt
\item We introduce a novel context-aware architecture that significantly improves sleep stage classification by integrating channel wise attention (squeeze and excitation) block with a residual network for precise EEG feature extraction. Coupled with multi-stacks of Bi-LSTM, the model captures complex temporal dependencies, setting a new benchmark for accuracy and MF1 score in sleep stage detection.

\item To the best of our knowledge, we present the first explainable deep-learning model for sleep staging, employing 1D-GradCAM to highlight important EEG features. This method improved the model's transparency and corresponded closely with the annotations of sleep experts (obtained in a blind validation experiment).

\item We present a data-efficient training method that optimizes the training process for increased efficiency while maintaining model robustness and achieving an 8x gain in training speed. 

\item Extensive experiments were performed on our model to demonstrate its superior performance compared to the existing state-of-the-art methodologies.

\end{enumerate}  
\section{Related Works}

\begin{figure*}
	\centering
		\includegraphics[scale=.065]{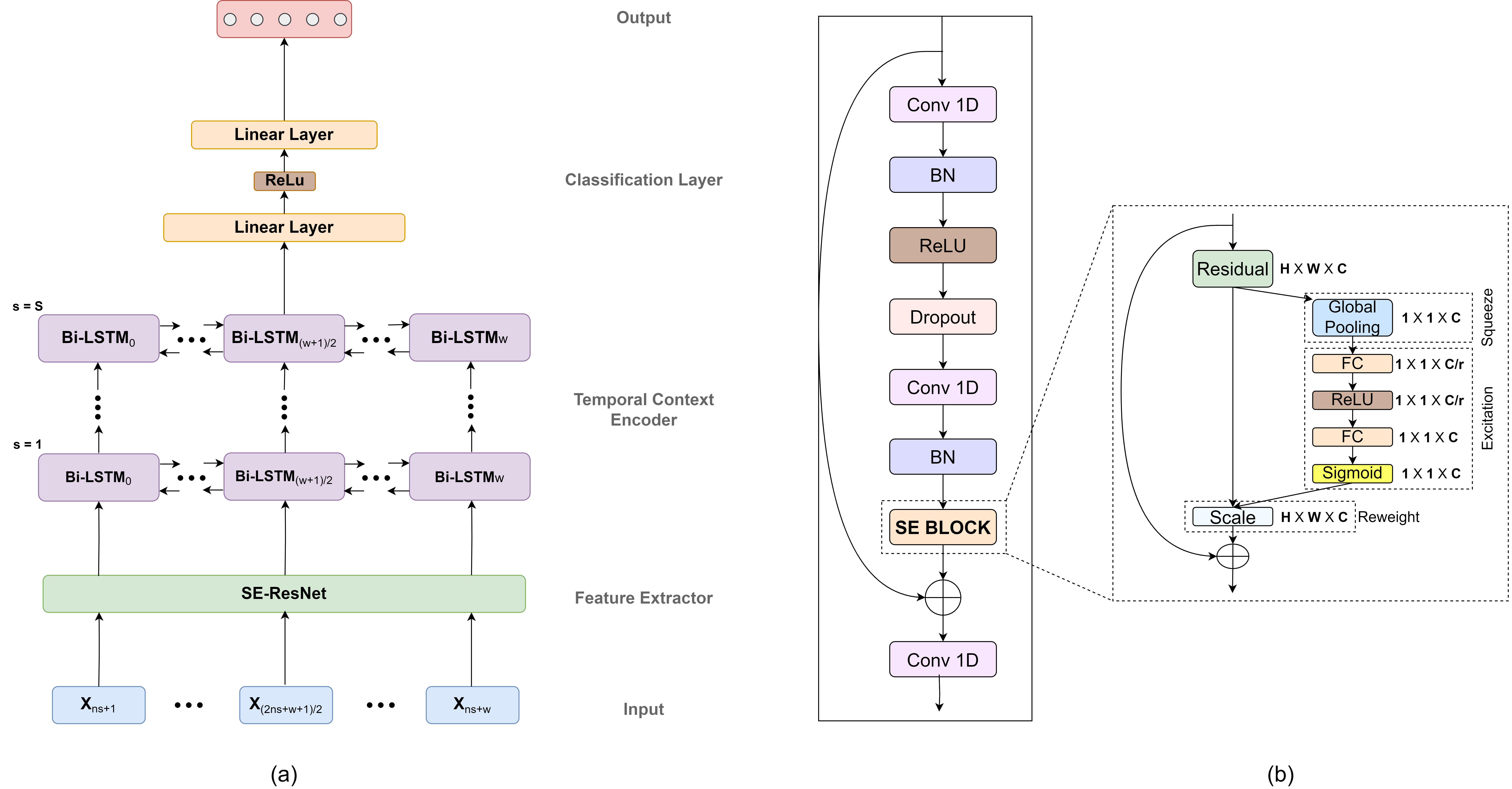}
	\caption{(a) Proposed Model Architecture (b) Description of SE block in SE-Resnet.}
	\label{FIG:4}
\end{figure*}

The initial studies centered on the extraction of manually designed features from electroencephalogram (EEG) signals and their application in classification tasks using conventional machine learning methods \cite{65,66}. Various features, such as spectrum
entropy, power spectral density, and statistical metrics, were derived from the electroencephalogram (EEG) signals. The classification was performed using algorithms such as Artificial Neural Networks (ANN) \cite{12}, Support Vector Machines (SVM) \cite{14,15,37}, Bootstrap aggregation \cite{43} and Random Forests (RF) \cite{16,40}. The performance of these methods was limited, as their effectiveness was hindered by the dependence on features that were manually designed. CNNs have been widely employed in the field of deep learning, playing a crucial role in several applications \cite{22,24,25,26,59,61}. The efficacy of CNNs arises from their capacity to independently extract features from raw data, hence obviating the necessity for manually engineered features. Additionally, these methods may not adequately capture intricate temporal relationships. RNNs were initially proposed as a means to effectively model and capture the temporal dependencies seen in sleep data. The LSTM \cite{6, 7} networks, which are a variant of RNNs, have garnered significant attention and adoption mostly because of their remarkable capability to effectively manage and model long-range dependencies. The utilization of LSTM models by researchers in the context of sequential modeling of EEG epochs resulted in enhanced accuracy when compared to conventional approaches \cite{27, 28,29}. The networks acquired knowledge of the intrinsic patterns and transitions among various sleep stages, hence enabling enhanced precision in classification. 

In EEG signal analysis, attention-based enhancements in RNNs and or attention in general have been explored to accentuate informative segments, as detailed in studies~\cite{17, 19, 20, 21, 64}. These methodologies enable the network to variably distribute attention across input segments. Conventional features extracted from EEG signals are often combined with raw data in deep learning frameworks, as noted by~\cite{38}, while the fusion technique proposed by~\cite{39} amalgamates fine-grained features with extensive patterns, yielding enriched feature representations and better classification accuracy. Approaches, such as the two-stage neural network proposed by Sun et al.\cite{31} that combines learned and manual features, and the integration of Bi-LSTM with attention mechanisms by Mousavi et al.\cite{34}, have shown promise in improving performance. Additionally, Phan et al.~\cite{44} emphasized learning from both raw signals and time-frequency images, adjusting learning rates for an equilibrium between generalization and overfitting, while also employing gradient blending to boost feature representation.

In brief, the progression of autonomous sleep stage categorization has shifted from the incorporation of manually designed features and conventional machine learning algorithms like SVM, RF, ANNs to the adoption of deep learning methodologies such as LSTM networks, attention mechanisms, and hybrid methodologies. The implementation of these techniques has resulted in a collective enhancement in accuracy, robustness, and capacity to manage a wide range of sleep patterns and disorders.

Despite these advancements, while many studies have used transformers and LSTM for their proficiency at modeling time series data, they frequently underperform in the feature extraction domain, especially in capturing the intricate frequencies and micro-architectures essential for delineating sleep stages. In our work, we put out focus towards the attention based feature extraction module. We have achieved superior results, as substantiated by our refined approach that prioritizes a more nuanced understanding and representation of EEG signals. This concerted effort not only highlights the importance of robust feature extraction in sleep stage classification but also sets a new precedent for future research in this domain.


\section{Methodology}
\subsection{Overview}
In our research study, we present a novel approach for sleep stage classification by using two advanced deep learning architectures: SE-ResNet and BiLSTM displayed in Figure \ref{FIG:4}. 

\subsubsection{SE-Resnet}

ResNet~\cite{45}, renowned for its performance on datasets like ImageNet~\cite{46}, is a deep neural network architecture that tackles the vanishing gradient problem with skip connection in residual blocks. Due to EEG's variability and complexity, we transitioned from ResNet to SE-ResNet for its superior dynamic feature capture and attention-based recalibration.

The Squeeze-and-Excitation Residual Network (SE-Resnet)~\cite{10} architecture features a Squeeze-and-Excitation (SE) block in each residual block. The SE block is intended to capture channel-wise dependencies by adaptively recalibrating the feature maps. Each residual block in SE-ResNet is made up of two or three convolutional layers, followed by a SE block and a shortcut connection. The SE block is made up of two major operations: squeeze and excitation. The global information from all channels is collected to generate a channel descriptor vector during the squeeze procedure. The channel descriptor vector is utilized to selectively re-weight the channels during the excitation procedure. This aids in emphasizing the critical channels while suppressing the less important ones, hence increasing the network's representational strength. When compared to ResNet, the addition of the SE block in ResNet allows for greater feature learning and improved performance in picture classification tasks.

\subsubsection{LSTM}

Long Short-Term Memory (LSTM) networks are a type of recurrent neural network designed to overcome the vanishing gradient problem in standard RNNs, which hinders learning long-term dependencies. LSTMs use a gating mechanism with three key parts: input, forget, and output gates, to control the flow and retention of information over time. This structure allows selective memory and forgetting of prior information, facilitating the learning of long sequences. Bidirectional LSTM (Bi-LSTM) extends this by processing data in both directions, enhancing context understanding, making it effective for signal processing tasks like speech recognition and NLP tasks like language translation. LSTMs have outperformed traditional RNNs and feedforward networks in various applications due to their ability to capture complex, long-term sequential patterns.

Let's assume that the input to the cell at time step t is denoted by \emph{x(t)}, and the output of the cell at time step t is denoted by \emph{h(t)}. Then, the equations for the LSTM cell can be written as follows:

\begin{gather}
f(t) = \sigma(W_f \cdot [h_{t-1}, x_t] + b_f) \\
i(t) = \sigma(W_i \cdot [h_{t-1}, x_t] + b_i) \\
\hat{c}(t) = \tanh(W_c \cdot [h_{t-1}, x_t] + b_c) \\
c(t) = f(t) \cdot c_{t-1} + i(t) \cdot \hat{c}(t) \\
o(t) = \sigma(W_o \cdot [h_{t-1}, x_t] + b_o) \\
h(t) = o(t) \cdot \tanh(c(t))
\end{gather}

where \(\sigma\) represents sigmoid activation function, \({W_f}\), \({W_i}\), \({W_c}\), \({W_o}\) are learnable weight matrices and \({b_f}\), \({b_i}\), \({b_c}\), \({b_o}\) are biases for the respective gates. The forget gate \emph{f(t)} determines how much of the previous cell state \emph{c(t-1)} should be retained. The input gate \emph{i(t)} determines how much of the new input \emph{x(t)} should be added to the cell state. The cell candidate $\hat{c}(t)$ is a new value that could be added to the cell state. The current cell state \emph{c(t)} is a combination of the previous cell state \emph{c(t-1)} and the new cell candidate $\hat{c}(t)$, weighted by the forget gate \emph{f(t)} and the input gate \emph{i(t)}. The output gate \emph{o(t)} determines how much of the current cell state \emph{c(t)} should be output as the cell output \emph{h(t)}.

The equations for Bi-LSTM are similar to those of LSTM, but with two separate LSTM cells, one processing the input sequence in the forward direction and the other processing it in the backward direction. Let's denote the forward hidden states as $\overrightarrow{h_1}, \overrightarrow{h_2}, \ldots, \overrightarrow{h_n}$ and the backward hidden states as $\overleftarrow{h_1}, \overleftarrow{h_2}, \ldots, \overleftarrow{h_n}$.  The output of the Bi-LSTM for each time step $i$ is $H_i = [\overrightarrow{h_i}; \overleftarrow{h_i}]$, where `;` denotes concatenation. The final output of the Bi-LSTM network is obtained by concatenating the outputs from the forward and backward directions from the middle cell (more details in Section~\ref{sec:contextInput}). \\

\subsubsection{Window and Stride}

\begin{figure*}[t]
	\centering
		\includegraphics[scale=.13]{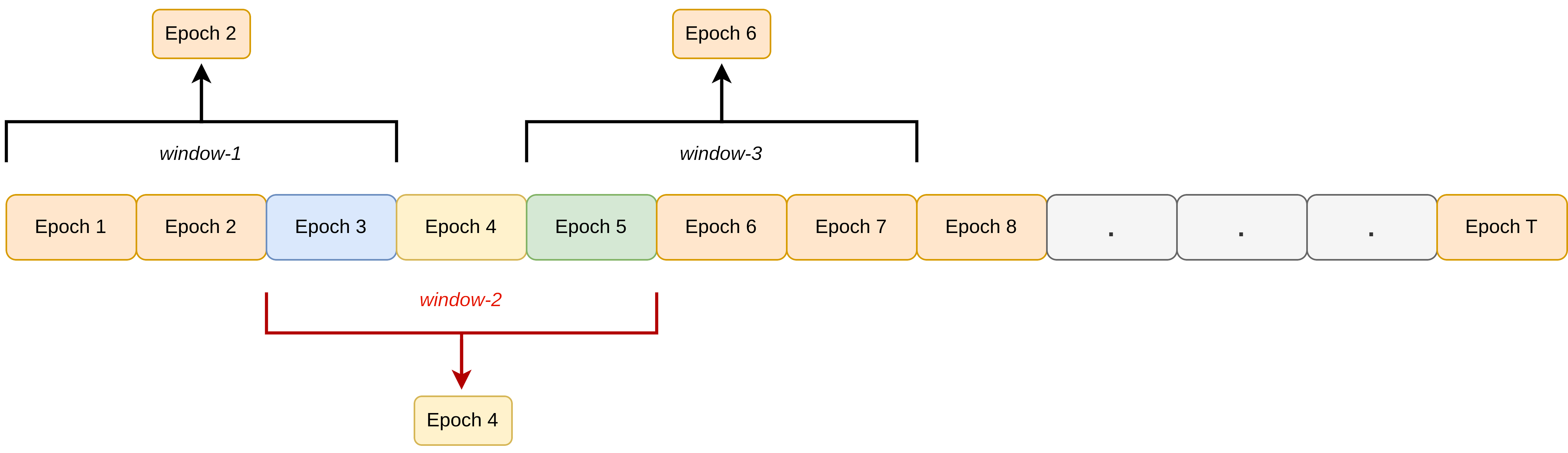}
	\caption{Notion of window and stride when the window size is 3 and stride is 2.}
	\label{FIG:3}
\end{figure*}

In sleep studies, an 8-hour patient recording is segmented into 30-second epochs. These epochs are sequentially analyzed using a sliding window approach. Each window contains a series of consecutive epochs acting as input for the model, defined by the "window size." and "stride" parameter dictates the window's shift after each step, affecting the overlap between successive frames. In Figure~\ref{FIG:3}, the window spans 3 epochs or 1.5 minutes.  A stride of 2 moves the window by two epochs (1 minute) each time, while a stride of 1 shifts it by one epoch (30 seconds). Thus, the model, using window-1 encompassing the first three epochs, predicts the label of epoch-2 while using epoch-1 and epoch-3 as context. This method adjusts overlap and data coverage by varying the stride. \\

\subsubsection{Explainable artificial intelligence: 1D-Grad-CAM}
Grad-Class Activation Mapping (CAM), introduced by Batra et. al.~\cite{11}, is a technique in Explainable AI (XAI) designed to identify critical regions in images that significantly affect the predictions of CNNs during image classification. In our work, we modified GradCAM, initially designed for 2D data such as images, to accommodate our 1D dataset. 

Grad-CAM calculates the gradients of the class output \(y^c\) with respect to the feature maps \(A^k\) of a specified convolutional layer. These gradients are globally averaged to obtain the weights \(\alpha^k_c\), signifying the importance of each feature map for the target class.

The formula for calculating these weights is given by:

\begin{equation}
\alpha^k_c = \frac{1}{V} \sum_i \frac{\partial y^c}{\partial A^k_{i}}
\end{equation}

where \(V\) is the total number of elements in the feature map. The GradCAM heatmap, \(L^c_{\text{GradCAM}}\), is then created by a weighted combination of these feature maps and applying a ReLU function to the result:

\begin{equation}
L^c_{\text{GradCAM}} = \text{ReLU}\left( \sum_k \alpha^k_c A^k \right)
\end{equation}

This process generates a heatmap aligned with the dimensions of the convolutional layer's feature maps, which is then normalized and resized for an integrated visualization with the original image.

In this specific study, the normalized and resized GradCAM heatmaps from the last layer of the Feature Extractor block were computed to enhance the detection of EEG signal patterns related to different sleep stages. This approach aims to provide a more detailed understanding of the factors influencing each sleep stage as determined by CNN.\\


\subsection{Mathematical Setup}
\subsubsection{Contextual Input}\label{sec:contextInput}

We propose a method that combines each epoch with its neighboring epochs, creating a contextual input that acknowledges the dependencies across time for more accurate classification.

In the analysis of sleep data, the fundamental unit of measurement is the sleep epoch, typically lasting 30 seconds. When examining a signal with a sampling rate of \( f_s \) Hz, each epoch contains \( 30 \times f_s \) data points. This temporal resolution is critical for capturing the detailed physiological variations associated with different sleep stages.

Input to the model, \( X_n \), is a windowed segment of the continuous signal. Specifically, for a given window size \( w \) $\{2w + 1 \mid w \in \mathbb{Z}\}$ and stride $g \in \{1, 2, \ldots, w\}$, the input is defined as:
\begin{equation}
X_n = [x_{ng+1}, x_{ng+2}, \ldots, x_{ng+w}] \quad \forall n \in \mathbb{Z} \cap [0, N-1]
\end{equation}
Each \( x \) represents an epoch and is a vector in \( \mathbb{R}^{1 \times 30f_s} \), while \( X_n \in \mathbb{R}^{w \times 30f_s} \). If \( T \) denotes the total number of epochs in a single subject recording, then \( N = \frac{T-w}{g} + 1 \). 

The output, \( Y_n \), corresponds to the sleep stage label of the central epoch in the input window. This is given by:


\begin{equation}
Y_n = \gamma_{\left(\frac{2ng + w + 1}{2}\right)}
\end{equation}


where $\gamma$ is the epoch label representing one of the five different sleep stages.

To construct the entire signal representation, \( R_{EEG} \), we concatenate the epochs:
\begin{equation}
R_{EEG} = \bigoplus_{i=0}^{T} x_i
\end{equation}
where \( \bigoplus \) denotes the concatenation operation.

This formulation allows for a structured approach to model wide temporal dynamics and relationships between consecutive epochs, capturing the progression and patterns inherent in sleep stages.\\

\subsubsection{Feature Extractor}
EEG signals are often complex and high-dimensional. Feature extraction helps the model make accurate predictions by identifying major variations in EEG data throughout sleep stages. Feature extraction is used to simplify and extract important data from this raw data. This stage carefully selects characteristics that capture sleep stage-specific patterns and attributes. These patterns encompass characteristics such as frequency and amplitude among other signal attributes. The features extrated in this layer are tailored to underscore crucial components while reducing noise in the raw data, thereby boosting accuracy by eliminating extraneous noise that could impede precise classification. This makes the machine learning model's classification task more robust. By incorporating the SE-ResNet block, which adaptively recalibrates feature responses, the model can effectively capture intricate EEG signal variations characteristic of different sleep stages.

In the context of our proposed architecture, we first transform the raw input sequence \( X_n \) to embedding \( E_n \) by $\mathit{F}: x \rightarrow e$ using SE-Resnet.


\begin{align}
E_n &= [\mathit{F}(x_{ng+1}), \mathit{F}(x_{ng+2}), \ldots, \mathit{F}(x_{ng+w})] \\
E_n &= [e_{ng+1}, e_{ng+2}, \ldots, e_{ng+w}] \quad \forall n \in \mathbb{Z} \cap [0, N-1]
\end{align}

\subsubsection{Temporal Context Encoder}
We improves the foundational feature extraction layer by adding a Bi-LSTM temporal context encoder. These signals are temporally dependent, unlike isolated EEG data points. Sleep stage classification requires understanding EEG signals' dynamic subtleties and transitions as one moves through sleep stages. This led us to use a Bi-LSTM module as a temporal context encoder, which detects detailed temporal patterns in EEG signals, representing sleep stage features, by embracing a greater temporal scope. The temporal context encoder's capacity to reveal fundamental links between sequential EEG segments makes it important, thus giving our model a deep understanding of these signals' evolution. The central epoch (as in Figure~\ref{FIG:4}(a)) has been selected for the output prediction, while the other epochs of the window function as providers of contextual information, integrating both past and future data into the model.

We define sequential processing of incoming data through Bi-LSTM layer, structured in a stacked format. Let the total number of these stacked Bi-LSTM layers be denoted by \( S \). Within this framework, the initial stack, indicated as \( s = 1 \), receives the embedding sequence \( E_n \). Subsequent stacks, labeled as \( s = 2, \ldots, S \), are designed to sequentially receive the output from the preceding stack, thus ensuring a deep, hierarchical representation of features.

At each time step \( i \), the output of any given stack \( s \) is represented by \( H_i^s \), which is a concatenation of the forward and backward hidden states, expressed as \( H_i^s = [\overrightarrow{h_i^s}; \overleftarrow{h_i^s}] \). The output of the final stack \( S \) was utilized. From this terminal stack, we specifically extract the middle cell's output, \( H_m^S \), to capture the surrounding context of the input sequence. This middle cell output is critical as it is presumed to hold the most significant temporal features, from both directions, given by \( H_m^S = [\overrightarrow{h_m^S}; \overleftarrow{h_m^S}] \).
\\

\begin{table*}[]
\centering
\caption{COMPREHENSIVE SUMMARY OF EMPLOYED DATASETS}
\label{tab:mergedTable}
\begin{tabular}{ccccccccccc} 
\hline
Dataset     & Subjects & W                                                       & N1                                                      & N2                                                       & N3                                                      & REM                                                     & Total Samples & EEG Channel Used & Scoring Method & k-folds \\ \hline
SleepEDF-20 & 20       & \begin{tabular}[c]{@{}c@{}}9118\\ 21.10\%\end{tabular}  & \begin{tabular}[c]{@{}c@{}}2804\\ 6.50\%\end{tabular}   & \begin{tabular}[c]{@{}c@{}}17799\\ 41.30\%\end{tabular}  & \begin{tabular}[c]{@{}c@{}}5703\\ 13.20\%\end{tabular}  & \begin{tabular}[c]{@{}c@{}}7717\\ 17.90\%\end{tabular}  & 43141         & Fpz-Cz           & R\&K           & 20         \\ 
SleepEDF-78 & 79       & \begin{tabular}[c]{@{}c@{}}66822\\ 34.00\%\end{tabular} & \begin{tabular}[c]{@{}c@{}}21522\\ 11.00\%\end{tabular} & \begin{tabular}[c]{@{}c@{}}69132\\ 35.20\%\end{tabular}  & \begin{tabular}[c]{@{}c@{}}13039\\ 6.60\%\end{tabular}  & \begin{tabular}[c]{@{}c@{}}25835\\ 13.20\%\end{tabular} & 196350        & Fpz-Cz           & R\&K           & 10         \\
SHHS        & 329      & \begin{tabular}[c]{@{}c@{}}43619\\ 14.30\%\end{tabular} & \begin{tabular}[c]{@{}c@{}}10304\\ 3.20\%\end{tabular}  & \begin{tabular}[c]{@{}c@{}}142125\\ 43.70\%\end{tabular} & \begin{tabular}[c]{@{}c@{}}60153\\ 18.50\%\end{tabular} & \begin{tabular}[c]{@{}c@{}}65953\\ 20.30\%\end{tabular} & 324854        & C4-A1            & R\&K           & 20         \\ \hline
\end{tabular}
\end{table*}

\begin{table*}[]
\centering
\caption{PERFORMANCE COMPARISON BETWEEN OUR MODEL AND PREVIOUS WORKS ON THE THREE DATABASES. WE MARK IN BOLD WHERE OUR MODEL PERFORMANCE IS BETTER OR EQUAL TO THAT OF ALL COMPATIBLE PRIOR WORKS}
\label{tab:comparison_table}
\begin{tabular}{|c|c|c|ccc|ccccc|}
\hline
\multirow{2}{*}{Database}    & \multirow{2}{*}{Model}    & \multirow{2}{*}{Fold} & \multicolumn{3}{c|}{Overall Metrices}         & \multicolumn{5}{c|}{Per1 Class F1-Score}                                                          \\
                             &                           &                       & ACC           & MFI           & $\kappa$             & W             & N1            & N2            & N3                                & REM           \\ \hline
\multirow{8}{*}{SleepEDF-20} & NAS ~\cite{52}              & 20                    & 82.7          & 75.9          & 0.76          & 86.2          & 38.8          & 87.8          & \multicolumn{1}{l}{\textbf{88.5}} & 77.5          \\
                             & IITNet ~\cite{53}          & 20                    & 83.6          & 76.5          & 0.77          & 87.1          & 39.2          & 87.7          & \multicolumn{1}{l}{87.7}          & \textbf{90.9} \\
                             & DeepSleepNet ~\cite{27}     & 20                    & 82.0          & 76.9          & 0.76          & 84.7          & 46.6          & 89.8          & \multicolumn{1}{l}{84.8}          & 82.4          \\
                             & DeepCNN ~\cite{57}          & 20                    & 83.2          & 78.1          & -             & 89.2          & 47.3          & 87.3          & \multicolumn{1}{l}{87.8}          & 83.9          \\
                             & XSleepNet2 ~\cite{44} & 20                    & 83.9          & 78.7          & 0.77          & -             & -             & -             & \multicolumn{1}{l}{-}             & -             \\
                             & SleepEEGNet ~\cite{34}      & 20                    & 84.3          & 79.8          & 0.79          & 89.2          & 52.2          & 86.8          & \multicolumn{1}{l}{85.1}          & 85.0          \\
                             & SleepContextNet ~\cite{54}  & 20                    & 84.8          & 79.8          & 0.79          & 89.6          & 50.5          & 88.4          & \multicolumn{1}{l}{88.5}          & 82.0          \\
                             & Ours                      & 20                    & \textbf{87.5} & \textbf{82.5} & \textbf{0.82} & \textbf{92.0} & \textbf{56.9} & \textbf{89.9} & \multicolumn{1}{l}{87.9}          & 85.9          \\ \hline
\multirow{9}{*}{SleepEDF-78} & NAS ~\cite{52}              & 10                    & 80.0          & 72.7          & 0.72          & 91.1          & 39.2          & 84.0          & 81.0                              & 68.1          \\
                             & SleepEEGNet ~\cite{34}      & 10                    & 80.0          & 73.6          & 0.73          & 91.7          & 44.1          & 82.5          & 73.5                              & 76.1          \\
                             & AttnSleep ~\cite{51}        & 10                    & 81.3          & 75.1          & 0.74          & 92.0          & 42.0          & 85.0          & \textbf{82.1}                     & 74.2          \\
                             & U-Time ~\cite{55}          & 10                    & -             & 76.0          & -             & 92.0          & \textbf{51.0} & 84.0          & 75.0                              & 80.0          \\
                             & TCNNNet ~\cite{55}          & 10                    & 82.5          & 76.1          & 0.76          & 92.4          & 48.1          &               & 75.0                              & 80.0          \\
                             & XSleepNet2 ~\cite{44}       & 10                    & 80.3          & 76.4          & 0.73          & -             & -             & -             & -                                 & -             \\
                             & FCNN+RNN ~\cite{51}         & 10                    & 82.8          & 76.6          & 0.76          & 92.5          & 47.3          & 85.0          & 79.2                              & 78.9          \\
                             & SleepContextNet ~\cite{54}  & 10                    & 82.7          & 77.2          & 0.76          & \textbf{92.8} & 49.0          & 84.8          & 80.6                              & 78.9          \\
                             & Ours                      & 10                    & \textbf{83.8} & \textbf{78.9} & \textbf{0.77} & 92.1          & 50.1          & \textbf{85.2} & \textbf{82.1}                     & \textbf{82.1} \\ \hline
\multirow{4}{*}{SHHS}        & AttnSleep ~\cite{51}        & 20                    & 84.2          & 73.5          & 0.78          & 86.7          & 33.2          & 87.1          & 87.1                              & 82.1          \\
                             & NAS ~\cite{52}              & 20                    & 81.9          & 75.3          & 0.74          & 84.8          & 36.9          & 84.4          & 80.9                              & 80.9          \\
                             & SleepContextNet ~\cite{54}  & 20                    & 86.4          & 80.5          & 0.81          & \textbf{89.2} & 52.0          & 87.6          & 83.3                              & \textbf{89.3} \\
                             & Ours                      & 20                    & \textbf{87.8} & \textbf{81.9} & \textbf{0.83} & 89.1          & \textbf{54.4} & \textbf{89.2} & \textbf{87.9}                     & 88.9          \\ \hline
\end{tabular}
\end{table*}

\subsubsection{Classification Layer}

For the effective integration of the central temporal features extracted from the last stack of the Bi-LSTM, \( H_m^S \), we utilize a series of \( k \) feed-forward neural networks (FNNs), where the output of one FNN serves as the input to the subsequent FNN in the sequence, to meticulously model detailed information. Each FNN encapsulates its own unique parameters: a weight matrix \( W_j \) and a bias vector \( b_j \) and \( j \in \mathbb{Z} \cap [0,  k -1] \). Upon receiving the input \( H_m^S \), consider \( j = 0 \), the initial FNN produces an output \( Z_0 \), calculated by executing a non-linear activation function, specifically, \text{ReLU}, on the linearly transformed input, formulated as :

\begin{equation}
Z_0  = \max\left((W_o \cdot H_m^S + b_o), 0\right)
\end{equation}

The final layer of the FNN produces a vector of size of number of classes (5), upon which a softmax function is subsequently applied, yielding a probability distribution across the classes.









\section{Results}
\subsection{Datasets}

\begin{figure}[t]
	\centering
		\includegraphics[scale=.045]{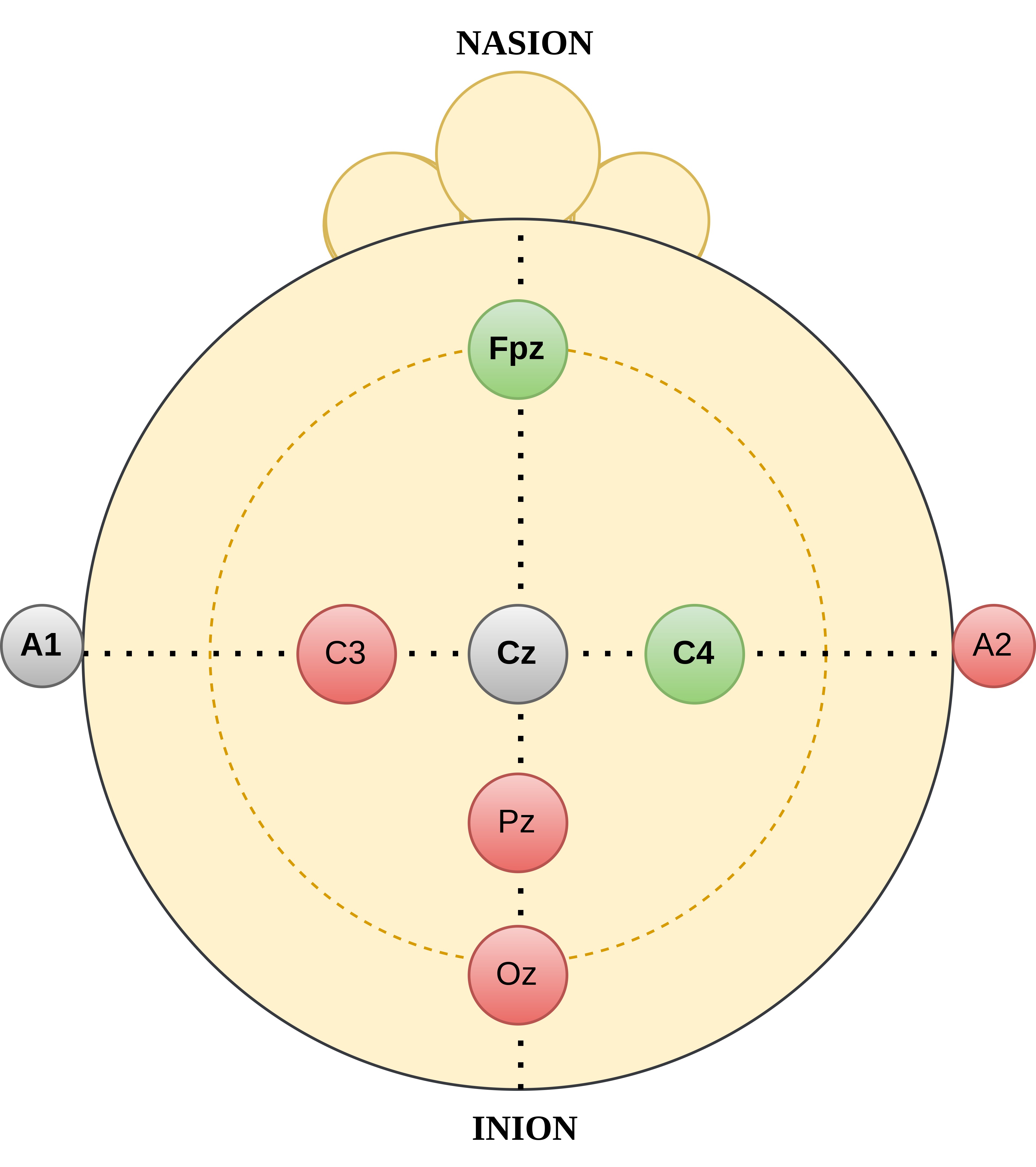}
	\caption{Representation of electrode placement of EEG. Green represents the electrode used, grey represents the used reference electrodes (in Table~\ref{tab:mergedTable}), Red represents unused electrodes for this study.}
	\label{FIG:1}
\end{figure}

We evaluated our model using C4-A1 EEG channel for SHHS \cite{49,50} and FPz-Cz EEG channel for  SleepEDF – 20 \cite{47,48} and SleepEDF-78 \cite{47,48} datset as mentioned in Table-\ref{tab:mergedTable}. The electrode placement can be seen in Figure~\ref{FIG:1} \\

\subsubsection{SleepEDF-20}
This dataset is a subset of the Sleep-EDF Expanded dataset, which is a popular benchmark dataset for determining sleep stages. The Sleep-EDF-20 dataset consists of 39 polysomnography (PSG) records taken from 20 patients (aged 25-34) as part of the Sleep Cassette project, which was conducted between 1987 and 1991 to investigate the effects of aging on sleep in healthy persons aged 25 to 101 years. Manual sleep classifying based on the Rechtschaffen and Kales (R\&K) standard criteria is provided in the dataset, with each 30-second epoch classified as W, N1, N2, N3, N4, REM, Movement (M), or UNKNOWN. The N3 and N4 stages were grouped together as N3, and the MOVEMENT and UNKNOWN categories were dropped.\\

\subsubsection{SleepEDF-78}
It is a part of the Sleep-EDF Expanded dataset that includes 78 patients' two consecutive day-night PSG recordings. One night recording was lost due to device failure in 3 participants, \(13^{th}\), \(36^{th}\), and \(52^{th}\) subject out of 78 subjects. The dataset uses the same inclusion/exclusion criteria and merges two sleep stages identical to the Sleep-EDF-20 dataset. Both datasets’ PSG recordings include two EEG channels (Fpz-Cz and Pz-Oz), one EOG channel, one chin EMG channel, and event markers, offering high-resolution data for sleep pattern analysis. Both EEG and EOG were sampled at 100Hz. Only Fpz-Cz channel of EEG from both SleepEDF-20 and SleepEDF-78 was used in our study.\\

\subsubsection{SHHS}
The Sleep Heart Health Study (SHHS) dataset consists of 6,441 subjects with a single full-night recording for each subject, collected over the course of two rounds of PSG recordings, Visit 1 (SHHS-1) and Visit 2 (SHHS-2). The manual scoring method described by R\&K was used for sleep scoring. In a move that mirrors the annotations found in the SleepEDF dataset, stages N3 and N4 have been collapsed into N3, while MOVEMENT and UNKNOWN were removed. Similar to \cite{51}, we used subjects with regular sleep and chose 329 participants from the SHHS database for our investigation. Single-channel EEG (C4-A1) was used for our study.

\begin{figure*}
    \centering
    \includegraphics[scale=.50]{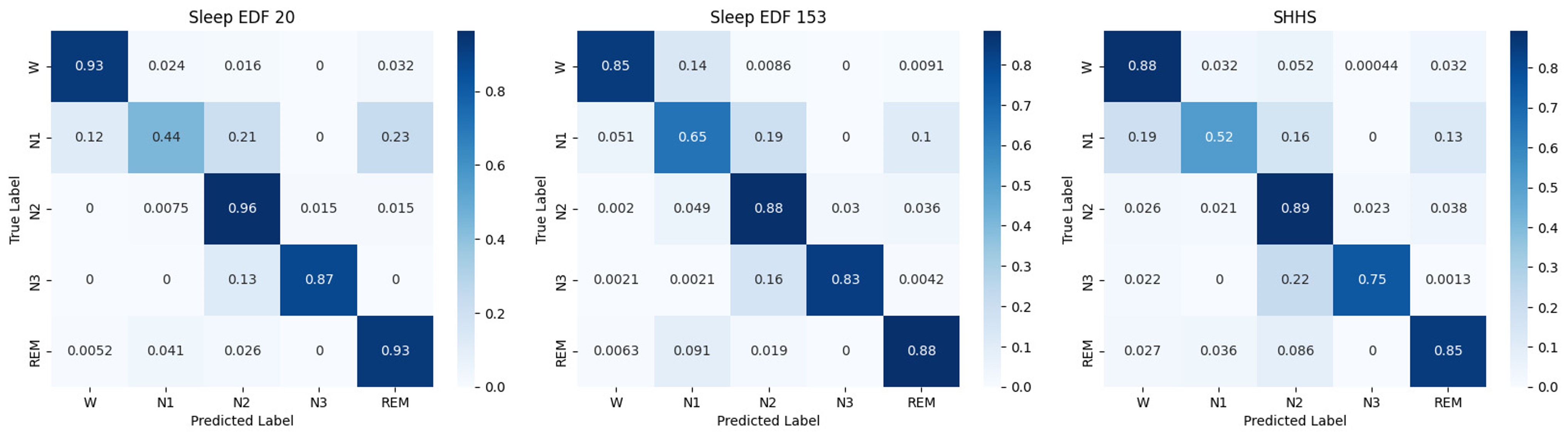}
    \captionsetup{justification=centering}
    \caption{Normalized Confusion matrices of the proposed model with window size-9 and stride-1 on the Fpz-Cz channel of SleepEDF-20 and SleepEDF-78 datasets and on C4-A1 channel of SHHS dataset, respectively. Each column in the confusion matrices represents predicted sleep stages, and each row represents the sleep stage annotations provided by experts. The dark color of the element on the diagonal represents better classification performance of respective categories.}
    \label{FIG:5}
\end{figure*}

\subsection{Performance Metrics}

In our evaluation, we used a robust set of metrics to thoroughly assess our proposed method's performance in automatically classifying sleep stages. We employed accuracy (ACC), kappa (\(\kappa\)), and F1-scores (F1) to gauge our model's proficiency. We examined both the individual F1-scores for each sleep stage and overall categorization effectiveness using the macro-averaged F1-score (MF1), providing a balanced evaluation across all sleep stages.



\subsection{Experimental Setup}
We conducted k-fold cross-validation (CV) tests to assess the performance and robustness of our model and ensure a fair comparison with earlier studies. These k-folds (as in Table~\ref{tab:mergedTable}) have frequently been included in research papers, facilitating a direct evaluation of our model's performance in comparison with previous studies.


In our study, we utilized PyTorch Lightning and trained our model on two Nvidia GeForce RTX 2080 Ti GPUs. We used the Adam optimizer with a learning rate of 0.001 to adaptively update model parameters, along with negative log-likelihood loss to refine classification of sleep data. The batch size was set at 128. Our model reached convergence within 45 training iterations across all datasets, indicating efficient training and satisfactory performance without excessive computational demand.

\subsection{Sleep Stage Scoring Performance}

Our model was trained and tested the SleepEDF-20, SleepEDF-78, and SHHS datasets over 20, 10, and 20 folds, respectively. As shown in Table \ref{tab:comparison_table}, the outcomes demonstrate the superiority of the proposed method across all performance metrics. Our model produced remarkable accuracy, MF1-score, Cohen's kappa, and class-wise F1-score results. Our model has achieved MF1 of 82.5\%, 78.9\%, 81.9\% and ACC of 87.5\%, 83.8\%, 87.8\% for SleepEDF-20, SleepEDF-78, and SHHS datasets, respectively.

In our study, we intended to evaluate the reliability and performance of our proposed method by conducting an exhaustive series of experiments on three distinct datasets: SleepEDF-20, SleepEDF-78, and SHHS. XSleepNet2~\cite{44}, IITNet~\cite{53}, DeepSleepNet~\cite{27}, SleepEEGNet~\cite{34}, SleepContextNet~\cite{54}, and NAS~\cite{52} were among the established methods with which we directly compared our algorithm. For consistency and equity, we utilized the same EEG channels for the proposed model as was used in the cited literatures.

\subsection{Ablation studies}
A series of ablation studies were methodically performed on the SleepEDF-20 dataset, for a total of 20 distinct folds. The methodology employed in this study entailed an in-depth study of the effects exerted by different components and features incorporated in our model. Through the systematic alteration of one variable, while maintaining the remaining parameters unchanged, our objective was to evaluate the distinct impacts of these elements on the performance of the model. Our goal was to identify the most influential factors that affect the model's effectiveness in dealing with the SleepEDF-20 dataset. The 20-fold cross-validation strategy was employed to ensure the reliability and robustness of our findings.\\

\subsubsection{Selection of Window Size and Stride}
During the trials, we tried out various permutations of window widths and strides to locate the configuration that worked best with the model. The length of the input sequence that is delivered to the model is determined by the window size, while the stride dictates the size of the step that must be taken to move the window along the EEG signal. Exploring the numerous possible combinations serves the objective of capturing a broad spectrum of temporal durations and determining which attributes are meaningful at varying sizes.

\begin{figure}
	\centering
		\includegraphics[scale=.37]{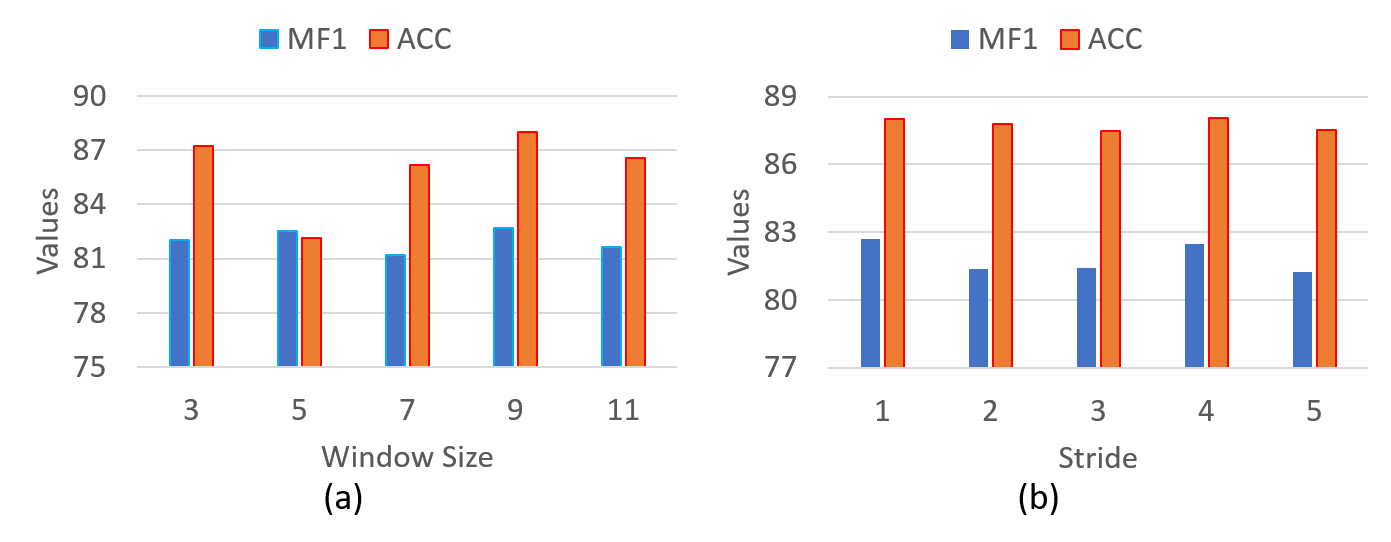}
	\caption{: Performance (MF1 and ACC values) of the model for different sizes of Window and Stride. (a) For Window sizes 3, 5, 7, 9 and 11. (b) For Stride varied from 1 to 5.}
	\label{FIG:6}
\end{figure}

Figure \ref{FIG:6} shows the performance of our model on SleepEDF-20 dataset when the size of the window and stride are varied. As can be seen, from Figure \ref{FIG:6}(a), the model has the highest classification performance e.g. MF1 and ACC are 82.68 and 88.02\% when window length is 9. From Figure \ref{FIG:6}(b), the model has the best  MF1 and ACC of 83.29\% and 88.02\% when Stride length is 1.

According to the findings, the optimal F1 score could be achieved with a window size of 9 and a stride of 4. This shows that adopting a bigger window size enables the model to catch a wider temporal context, which could potentially allow it to extract more informative features from the EEG signals. The model begins to lose relevant information and a possible information bottleneck develops when the number of Bi-LSTM layers exceeds the optimum level 9 in our case. As additional layers are added, the model's performance therefore begins to suffer, which results in decreasing returns. Adopting an optimally large window size also enables the model to acquire more data at once. In addition, the increased stride of 4 enables more efficient processing by cutting down on redundant computations while still keeping a suitable overlap between neighboring windows. \\


\subsubsection{Selection of stacks of Bi-LSTMs and SE-ResNet}
\begin{figure}[t]
	\centering
		\includegraphics[scale=.36]{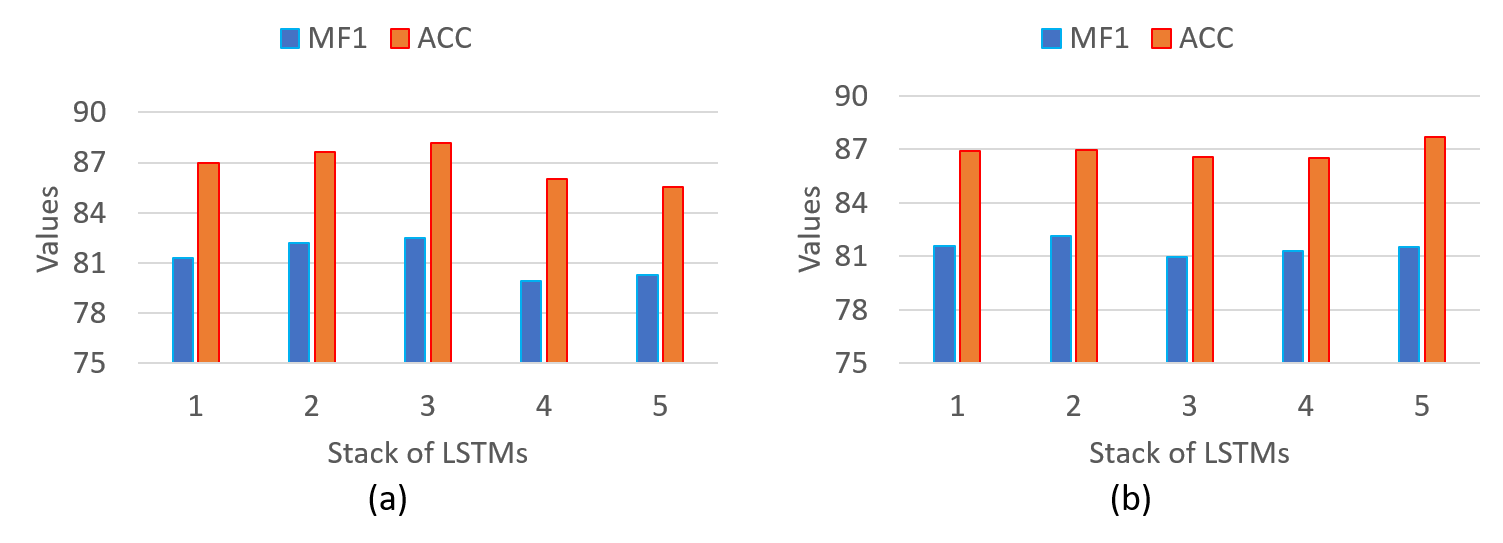}
	\caption{Performance metrics (MF1 and ACC) of model for varying number of LSTM stacks from 1 to 5. (a) For SE-Resnet-18 as feature extractor (b) For SE-Resnet-34 as feature extractor}
	\label{FIG:7}
\end{figure}

While selecting the feature extractor , we observed that SE-Resnet outperformed Resnet as the feature extractor. This higher performance is due to the inclusion of the SE module, which provides channel-specific attention and improves feature extraction quality. It was observerd that SE-ResNet-18 performed better in comparison to SE-ResNet-34 (Figure~\ref{FIG:7}). This is because  SE-ResNet-34, a deeper model with more layers have overfitted the data.

This work also looked into the effect of changing the total number of Bi-LSTM stacks contained within the model, which was another facet that was explored. The focus of this study was to determine the optimum number of stacks for the Bi-LSTM by putting it through a series of performance evaluations with varying configurations of stacks (the number of stacks tested ranged from one to five). The investigation found that the arrangement with three stacks of Bi-LSTM achieved the greatest possible F1 score. 

Figure \ref{FIG:7} presents the comparative performance metrics of our model on the SleepEDF-20 dataset, exploring various configurations of Bi-LSTM stacks paired with different variants of SE-Resnet as feature extractor. As depicted in Figure \ref{FIG:7}(a), the model with SE-Resnet-18 architecture achieved best classification results, attaining an MF1 score of 82.5\% when integrated with 3 Bi-LSTM stacks. Conversely, as shown in Figure \ref{FIG:7}(b), while the model with SE-Resnet-34 as a feature extractor reached its peak performance with 2 Bi-LSTM stacks, yielding an MF1 score of 82.29\%, it failed to surpass the efficacy of the SE-Resnet-18-based model.

The enhancement in Bi-LSTM layers serves to refine the model's capability in capturing increasingly complex temporal associations and in extracting sophisticated attributes from EEG data, pivotal for sleep stage classification. 

The inclusion of a multiple Bi-LSTM stacks have increased the model's capacity to accurately reflect the complex temporal dynamics inherent in EEG signals, thereby boosting its performance. Beyond certain point, additional stacks contribute to model complexity without proportionate gains in performance, potentially restricting the model's generalizability. This limitation arises from the additional layers' propensity to capture noise or training-specific patterns, coupled with escalated computational costs, all of which bear implications for the model's practical deployment in clinical settings.\\

\subsubsection{Efficient Training Approach}
During experimentation, we initially trained our model with an optimal window size of 9 while varying the stride. However, during testing, we used data with a fixed stride of 1 to ensure the clinical applicability by predicting label for all the epochs in a recording as well as to compare our results with other baselines. Surprisingly, we found that by training the model with the same window size of 9 but a larger stride of 4, we achieved comparable performance. This modification accelerated the training process, achieving an 8x speedup compared to the original stride of 1. This change in stride meant that we utilized only 25\% of the total available training epochs. Nevertheless, the model demonstrated an ability to generalize effectively despite the reduced training samples.

\begin{figure}
	\centering
		\includegraphics[scale=.38]{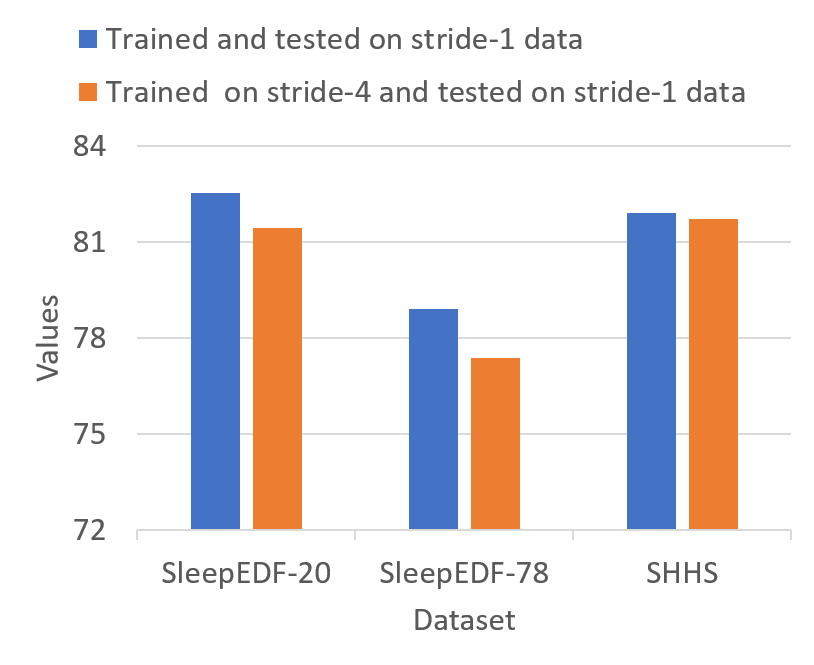}
	\caption{Performance of the model (MF1 values) when trained and tested with stride-1 vs when trained with stride-4 and tested on stride-1 on all the 3 datasets}
	\label{FIG:8}
\end{figure}

Figure~\ref{FIG:8} shows a comparison of the averaged MF1 scores obtained from our model trained with two different stride values, stride-1 and stride-4 while maintaining a fixed window size of 9 across all three datasets. In Table~\ref{tab:rapid_performance}, we present the per-class F1 scores for the model trained with a stride of 4. Utilizing all epochs in the datasets, with a stride of 1 during training, resulted in MF1 score improvements of only 1.3\%, 2\%, and 0.23\% across the SleepEDF-20, SleepEDF-78, and SHHS datasets when compared to training with reduced training samples (i.e when stride was kept 4). This transition to a larger stride offers improved computational efficiency by reducing redundancy in the data. Nevertheless, it is worth noting that this adjustment maintains an adequate level of overlap between windows, a critical factor for preserving important temporal relationships.


\section{Discussion}
This work introduces a comprehensive framework that seamlessly integrates feature learning capabilities and classification capacity, with a specific focus on incorporating temporal context. To enhance the interpretability of our model, we leverage GradCam and t-SNE visualization techniques, offering valuable insights into the decision-making processes of the model specially to clinicians.

\subsection{Selection of middle epoch of the window}
We extracted the output from the middle epoch of the window, yielding advantages for the model:\\

\textbf{Contextual Information}: A window's central epoch offers a fair depiction of the surroundings, as it merges both past and future information into the model. Enabling the model to comprehend the temporal context around the particular time point of interest more thoroughly. It ensures that the model can take into account necessary information from both earlier and later time steps, thereby enhancing its capacity to detect significant relationships and patterns in the EEG signal.

\textbf{Reduction of Boundary Effects}: By selecting the middle epoch, the window's possible boundary effects are lessened compared to when just only single epoch is taken into account. The performance of the model may be adversely affected by singular epochs' potential contextual deficiencies or inadequate temporal patterns. Additionally, opting for the middle epoch also ensures that the model gains access to a well-balanced and representative temporal context, leading to more accurate predictions.

\begin{table}[]
\centering
\caption{CLASSIFICATION PERFORMANCE OF OUR MODEL WHEN TRAINED WITH STRIDE-4 AND TESTED WITH STRIDE-1}
\label{tab:rapid_performance}
\begin{tabular}{|c|ccccc|}
\hline
\multirow{2}{*}{Dataset} & \multicolumn{5}{c|}{Per class F1 Scores} \\
                         & W      & N1     & N2     & N3    & REM   \\ \hline
SleepEDF-20              & 91.28  & 55.20  & 88.29  & 89.54 & 85.84 \\
SleepEDF-78              & 91.83  & 49.73  & 84.79  & 79.31 & 81.17 \\
SHHS                     & 89.55  & 51.77  & 89.26  & 88.33 & 89.72 \\ \hline
\end{tabular}
\end{table}

\begin{figure*}[h]
	\centering
		\includegraphics[scale=.5]{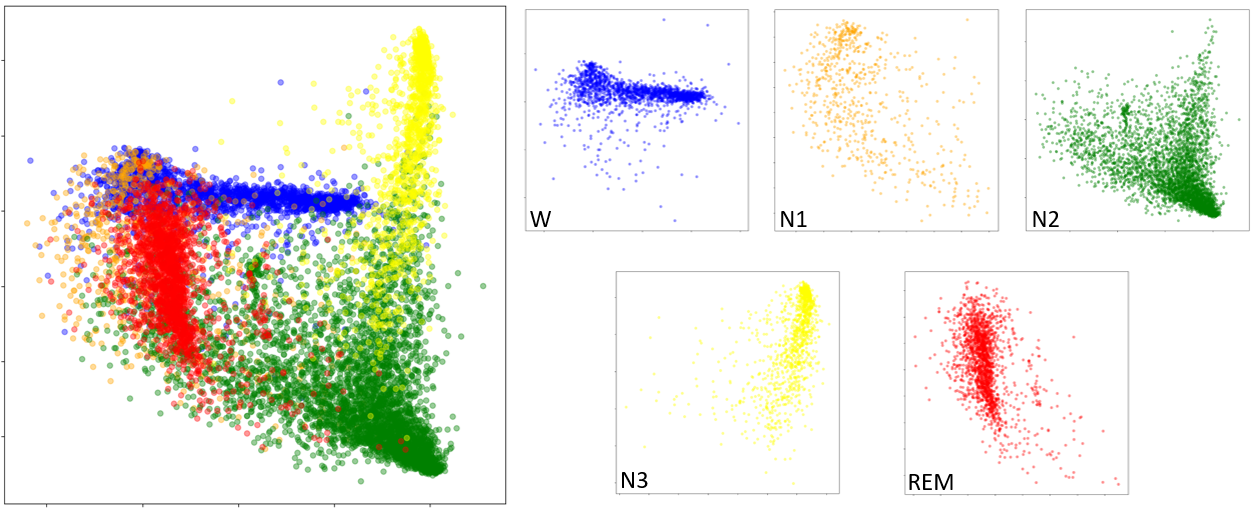}
	\caption{t-SNE plot of features derived from the last convolutional layer of the SE-Resnet }
	\label{FIG:9}
\end{figure*}

\subsection{Model Explainability}
\subsubsection{t-SNE}
The aforementioned results highlight the ability of our model to effectively extract vital features from raw EEG data. To visualize the high dimensional distribution of these feature we used the t-Distributed Stochastic Neighbour Embedding (t-SNE) to show the intricate features derived from the last convolutional layer of the SE-Resnet feature extractor, as seen in Figure \ref{FIG:9}. This approach enabled us to transform high-dimensional data into a two-dimensional representation to visualize a well-defined clustering pattern that closely corresponds to five separate sleep stage classes. 

It is worth noting that all classes, except for N1, have clear and discernible limits. It is crucial to highlight that there exists a substantial degree of overlap between the N1 class and both the REM and W classes. This finding suggests that the model has challenges in accurately distinguishing cases in the N1 class. This could be because of the transitional nature of N1 and the prevalence of alpha waves in both wakefulness and light sleep stages which present significant challenges in differentiation, exacerbated by external disturbances and restlessness that induce wake-like brain activity during N1. Concurrently, the presence of mixed frequency brain waves in both N1 and REM stages complicates their distinction, particularly in the absence of definitive REM markers like rapid eye movements and muscle atonia, further obfuscating accurate classification during ephemeral awakenings and sleep stage transitions. This assertion is further supported by the consistent misclassifications of the N1 sleep stage as either REM or W sleep stages, as evident in the corresponding confusion matrices presented in Figure \ref{FIG:5}. Moreover during early N3, where delta waves are just beginning to appear, can resemble late N2, especially if sleep spindles and K-complexes are still present. This can make it difficult to pinpoint the exact moment of transition between these stages. Since this transition from N2 to N3 is gradual, with an increasing proportion of delta waves, classification during this phase was challeanging for the model too (Figure \ref{FIG:5}). \\

\begin{figure*}[h]
	\centering
		\includegraphics[scale=.70]{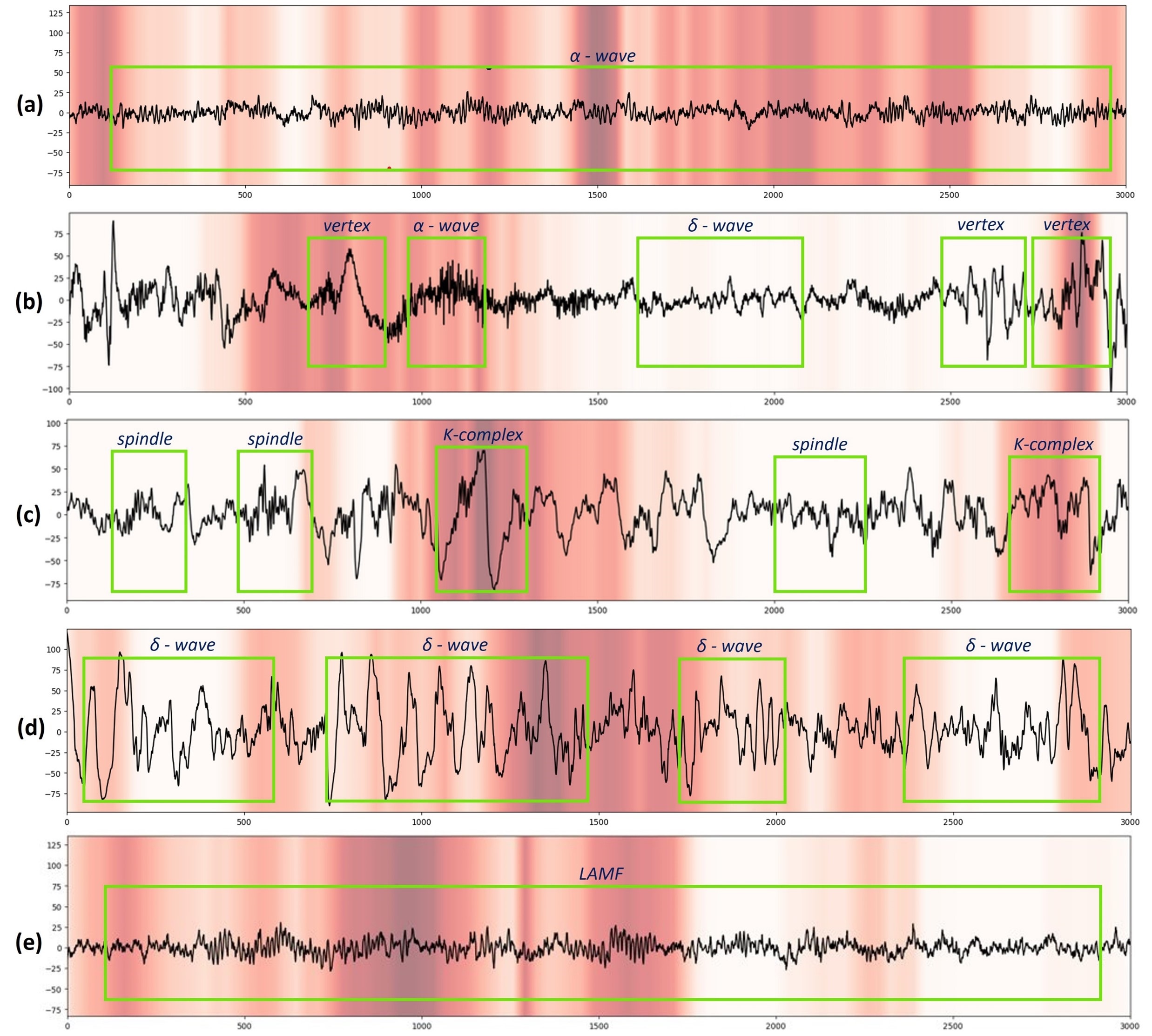}
	\caption{1D-GradCAM visualization of raw EEG epochs along with sleep micro-structures shown in green boxes. (a) Sleep stage - W and alpha wave, (b) Sleep stage - N1 and sleep vertices, alpha-wave and delta-wave (c) Sleep stage -N2 along with sleep spindle, K-complexes (d) Sleep stage- N3 with delta-waves(e) Sleep stage-REM and Low amplitude mixed frequency (LAMF) waveform}
	\label{FIG:10}
\end{figure*}

\subsubsection{GradCAM}
To enhance the interpretability of our model, especially for sleep specialists, we employed 1D GradCAM visualizations to elucidate the classification decisions across various sleep stages, as depicted in Figure \ref{FIG:10}. This technique employs progressively darker red hues to highlight specific segments of the input signal that significantly impact the model's final classification decision. In the actual clinical setting, during the manual scoring procedure performed by experts in sleep laboratories, a thorough examination is conducted on other channels of PSG, with particular emphasis on the EEG channel. The scoring conducted by professionals is based on the identification of multiple different artifacts (Table~\ref{tab:EEGactivity}) that are observable within the raw data. 

A cross-verification was performed to assess the classification performance of our model in distinguishing raw EEG data, with a single-blind evaluation of sleep specialists (co-authors SM and PRS from the Neurology Department, National Institute of Mental Health and Neurosciences (NIMHANS),  a tertiary hospital in India). The experts were presented with identical EEG data windows as our model, without any supplementary PSG modalities such as EOG or EMG. In order to come up with their predictions, the experts conducted an analysis of the specific regions demarcated by green boxes as depicted in Figure \ref{FIG:10}.

The experts carefully sought out distinctive micro-structures (in Table~\ref{tab:EEGactivity}) that are commonly associated with various sleep stages. Figure~\ref{FIG:10} shows the ability of our model to precisely detect different patterns seen in different sleep stages. Figure~\ref{FIG:10} (a) effectively recognizes alpha waves during the W stage. In Figure~\ref{FIG:10} (b) the model partially identifies the occurrences of sleep vertex and delta waves in sleep stage N1. Moreover, Figure~\ref{FIG:10} (c) demonstrates that the model prioritizes the identification of K-complex patterns, which serve as reliable markers for sleep stage N2. It effectively detects instances of delta waves during sleep stage N3, as seen in Figure~\ref{FIG:10} (d). Finally, Figure~\ref{FIG:10} (e) illustrates a robust performance in the classification of Rapid Eye Movement (REM) sleep phases.

The model demonstrates significant efficiency in the identification of sleep vertex and delta waves during sleep stage N1, which are critical microstructures for the classification of an epoch as N1. This capability majorly arises from SEResnet's ability to isolate distinctive features from these specific regions within the raw EEG signal. 


Our model demonstrated a high level of proficiency in capturing these sleep microstructures. It demonstrates a significant correspondence between the focus of our model and the specific regions within the raw EEG signal that play a crucial role in classifying sleep stages. In addition, this alignment closely resembles the approaches employed by sleep experts. The finding demonstrates the model's ability to prioritize similar signal segments that professionals seem to depend on for precise sleep epoch classification.

\section{Conclusion}
This study presents a novel end-to-end deep learning framework for the classification of sleep stages using single-channel EEG data. The proposed model with model with SE-Resnet as a Feature Extractor and, stacked Bi-LSTM as a Temporal Context Encoder exhibited commendable performance on three publicly available datasets. By utilizing the novel use of 1D-GradCAM visualization, our method offers a means to get insight into the decision-making mechanism of the model as evidenced by the close alignment between the model outcomes and expert annotations. Our model achieved superior results compared to the existing baseline methods, thereby setting a new benchmark for subtype classification performance on the SleepEDF and SHHS datasets. The integration of an accelerated training methodology saved extensive computation without compromising on performance. This framework could also be adapted to other PSG modalities typically used for sleep staging. Our future work involves enhancing the detection of the N1 sleep stage with a major focus on feature extractor, which remains an ongoing challenge in the field.


\section*{Acknowledgment}

We acknowledge financial assistance from IHub-Data, a Technology Innovation Hub (TIH) funded by Department of Science \& Technology (DST), Government of India (GoI) under the National Mission on Interdisciplinary Cyber-Physical Systems (NM-ICPS), International Institute of Information Technology Hyderabad (H2-003).

\section*{References}

\bibliographystyle{unsrt}

\bibliography{cas-refs}

\end{document}